\documentclass[aps,superscriptaddress,nofootinbib,showpacs]{revtex4}\usepackage{amssymb}
\usepackage{amsmath}
\usepackage{amsthm}
\usepackage{graphicx}
\usepackage{graphics}

\newcommand\scri{\mathcal{I}}

\newtheorem{thm}{Theorem}

\newtheorem{defn}{Definition}
\theoremstyle{remark}

\begin{document}

\title{A simple proof of Birkhoff's theorem for cosmological constant}
\author{Kristin Schleich}
\author{Donald M. Witt}
\affiliation{Department of Physics and Astronomy, University of British Columbia,
Vancouver, British Columbia \ V6T 1Z1}
\affiliation{Perimeter Institute for Theoretical Physics, 31 Caroline Street North, Waterloo,
ON, N2L 2Y5, Canada}
\date{\today}

\begin{abstract}
We provide a simple, unified proof of Birkhoff's  theorem for the vacuum and cosmological constant case, emphasizing its local nature. We discuss its implications  for the maximal analytic extensions of Schwarzschild,  Schwarzschild(-anti)-de Sitter  and Nariai spacetimes. In particular, we note that the maximal analytic extensions of extremal and over-extremal Schwarzschild-de Sitter spacetimes exhibit no static region. Hence the common belief that Birkhoff's theorem implies staticity is false for the case of positive cosmological constant. Instead, the correct point of view is that  generalized Birkhoff's theorems  are local uniqueness theorems whose corollary is that locally spherically symmetric solutions of Einstein's equations  exhibit an additional local Killing vector field.
\end{abstract}
\maketitle

Birkhoff's theorem, that any spherically symmetric vacuum solution of Einstein's equations is locally isometric to a region in Schwarzschild spacetime,  is not only a classic contribution to general relativity but is also an important tool in gravitational physics and cosmology.
First proven for the vacuum Einstein equations, this theorem is
 traditionally
credited to  Birkhoff \cite{Birkhoff}; however it was recently rediscovered by Deser, when providing an alternate proof, that  this theorem was published two years earlier by Jebsen \cite{Jebsen,Deser:2004gi,VojeJohansen:2005nd}. (An english translation of Jebsen's paper with an introduction by Deser  was published in General Relativity and Gravitation in 2005 \cite{Jebsen:2005,Deser:2005xx}.) In older references,  this theorem is credited as independently discovered by not only Birkhoff and Jebsen but also Alexandrow \cite{Alexandrow} and Eiesland \cite{Eiesland1925}.\footnote{Birkhoff's theorem is one of the corollaries to a theorem of Eiesland, which applies to Einstein spaces, not just the vacuum case  \cite{Eiesland1925}. Though published in 1925, this result was communicated to the American Mathematical Society on March 26, 1921 \cite{Eiesland1921}. }  The multiple provenance of this theorem is a testament to its  significance in general relativity. It is natural to consider its extension to other situations in gravitational physics. Birkhoff's theorem readily generalizes to the inclusion of electromagnetic fields \cite{Hoffmann}, cosmological
constant \cite{Eiesland1925}, with explicit exhibition of both the Schwarzschild-de Sitter and Nariai solutions in \cite{ MorrowJones:1988yw,MorrowJones:1993zu} (also in \cite{rindler} without citation of previous references), and to other symmetry groups \cite{Taub:1950ez,Goenner1970}.\footnote{Section 3 of the paper by Goenner \cite{Goenner1970}  provides a historical account of this theorem and its various generalizations to 1970.}  Generalized Birkhoff's theorems have also been proven in  lower and higher dimensions \cite{Kiem:1993zd,Schmidt:1997mq,Severa:2002tf,Bronnikov:1994ja,AyonBeato:2004if,Keresztes:2007vv}, certain alternate theories of gravity \cite{Reddy:1973,Krori1977,Ramaswamy1979,Neville:1979fk,Rauch:1981tva,Riegert:1984zz,Wiltshire:1985us} including Lovelock gravity \cite{Whitt:1988ax,Zegers:2005vx,Deser:2005gr,Maeda:2007uu}, and shown not to hold on Randall-Sundrum branes \cite{Bruni:2001fd} and in some modified theories of gravity \cite{Moffat:2007by,Dai:2008zza}.

Although Birkhoff's theorem is  a classic result, many current textbooks and review articles on general relativity no longer provide a proof or even a careful statement of the theorem.  Frequently it is cited as proving that the spherically symmetric vacuum solution is static. This is clearly not the case as recognized in many (but not all) proofs of the theorem (See, for example, the comment on Jebsen's proof in \cite{Ehlers:2006xx}). The interior region of  the maximal analytic extension of Schwarzschild spacetime is not static; instead, it exhibits a spacelike Killing vector field. Of course, other regions of this spacetime do exhibit a locally static Killing vector field. However, as pointed out in this paper, even this somewhat generous interpretation of the folklore of staticity is no longer true for the generalization of Birkhoff's theorem to positive cosmological constant; the maximal analytic extensions of the extremal and over-extremal Schwarzschild-de Sitter spacetimes have no static region anywhere, even though they still exhibit an additional Killing vector field.  Hence, the correct viewpoint on Birkhoff's theorem and its generalizations is that they are local uniqueness theorems; they prove that locally spherically symmetric solutions are locally isometric to a region in a specified set of parameterized spacetimes.  Furthermore, their corollary is that locally spherically symmetric solutions  exhibit an additional local Killing vector field; however this field is not necessarily timelike. Proper  statement of generalized Birkhoff's theorems  and their corollary can be important for correct application to physical situations, especially cosmological ones and those involving alternate theories of gravity.

In this paper, we will provide a simple proof of Birkhoff's theorem, emphasizing its local nature. This proof addresses not only the vacuum case but also the  positive and negative cosmological constant case in a unified form. It utilizes a particularly convenient gauge choice that removes the need to address the timelike, spacelike and null foliations separately as done,for example, in \cite{hebirk}. Furthermore, the resulting form of the solution unifies the timelike and spacelike cases. We then discuss its manifestation in the maximal analytic extensions of Schwarzschild, Schwarzschild-anti-de Sitter and Schwarzschild-de Sitter spacetimes and how the extremal and over-extremal Schwarzschild-de Sitter spacetimes illustrate the above points.

It is useful to begin by recalling that a smooth manifold is a metrizable space that is locally the same as ${\mathbb R}^n$. Precisely,
\begin{defn} A
metrizable space $M^n$ is a  smooth manifold if:  

(i) Every point has a neighborhood
$U_\alpha$ which is homeomorphic to an open subset of ${\mathbb R}^n$ via a
mapping $\phi_\alpha:U_\alpha \to {\mathbb R}^n$.

(ii) Given any two
neighborhoods $U_\alpha$, $ U_\beta$ with nonempty intersection,  the mapping 
\begin{equation*} \phi_\beta
\phi_\alpha^{-1}: \phi_\alpha(U_\alpha \cap U_\beta) \to
\phi_\beta(U_\alpha \cap U_\beta)
\end{equation*} is a smooth mapping between subsets
of ${\mathbb R}^n$.
\end{defn}
\noindent A neighborhood $U_\alpha$ with its associated $\phi_\alpha$ is termed a chart. An atlas is the collection of smooth charts, $\{  U_\alpha, \phi_\alpha\}$,  needed to cover $M^n$. In general, more than one chart is needed to cover a smooth manifold. A smooth metric $g_{ab}$ is a smooth symmetric nondegenerate covariant $2$-tensor on $M^n$. In a given chart, this metric can be written explicitly in terms of a line element
\begin{equation} ds^2 = g_{\alpha\beta}dx^\alpha dx^\beta \ .\end{equation}
As more than one chart may  be used to cover $M^n$, this expression is local; however,  the metric itself is global.

Open neighborhoods appear in a natural way in the definition of  locally symmetric tensor fields:

\begin{defn}\label{lss1} A tensor field, $T_{ab\cdots }{}^{cd\cdots }$,
on a manifold $M^n$ is {\it locally spherically symmetric}  if and only if
every point in $M^n$ has an open neighborhood $U$ such that the following 
conditions are satisfied:

(i) There is a set of three independent vector fields $\{\xi _i{}\}$, $i= 1, 2,3$ on $U$ which 
generate a faithful representation of the Lie algebra of $SO(3)$ and  the orbits of the vector fields are spacelike and
two dimensional.

(ii) ${\mathcal L} _{\xi _i}T_{ab\cdots }{}^{cd\cdots } \bigg| _U = 0$ for these 
vector fields. 
\end{defn}

\noindent Note that the orbits of $\{\xi _i{}\}$ need not be contained entirely in $U$; in fact they are not for neighborhoods on the round 2-sphere  discussed below.

A key example of such a tensor is a locally spherically symmetric metric $g_{ab}$ on $M^n$. For this case,  the
vector fields $\{\xi _i{}\}$ are {\it local Killing vector fields}, that is they are Killing vector fields in the neighborhood $U$. If these vector fields, given in one neighborhood $U$, can be smoothly defined on all of $M^n$ such that Definition \ref{lss1} is satisfied by this same set everywhere, then they are {\it global Killing vector fields}. 
\begin{defn} A manifold $M^n$ with locally spherically symmetric metric $g_{ab}$ is a locally spherically symmetric space. 
\end{defn}
If, in addition, the local Killing vector fields can be extended to global Killing vector fields, the space is {\it globally spherically symmetric}. In this case,  the orbits of the Killing vectors are necessarily closed 2-manifolds.

The canonical example of a globally spherically symmetric space is $S^2$ with round metric. This space can be realized as the set of points a unit distance from the origin in Euclidean space, ${\mathbb R}^3$, $S^2=\{(x,y,z)\in { \mathbb R}^3| x^2 + y^2+z^2=1\}$. The induced metric from this embedding is written
\begin{equation} d\Omega_2^2 = d\theta^2 +\sin^2 \theta d\phi^2\label{spherical}\end{equation}
in spherical coordinates with range $0<\theta<\pi$, $0<\phi<2\pi$.
Note that, strictly speaking,  (\ref{spherical}) is the metric as given in one of a set of charts needed to cover the sphere; this chart covers all of the 2-sphere except for an arc connecting the north and south poles.  Additional charts need to be chosen such that the entire 2-sphere is covered. In the presence of spherical symmetry, each can be taken such that the metric  is again of the form (\ref{spherical}).  Hence use of (\ref{spherical})  in a chart can be taken to imply such a choice of charts on $S^2$.

It is well known that rotations about the origin of  ${\mathbb R}^3$ are elements of $SO(3)$ and are generated by a set of infinitesimal $3\times 3$  rotation matrices $O_i$ with algebra $[O_i,O_j] = \epsilon_{ijk}O_k$. These rotations leave the sphere with its induced metric invariant. The infinitesimal generators of these rotations correspond to Killing vector fields on the sphere; for example, in the spherical coordinates of (\ref{spherical}) these can be taken to be
\begin{equation}\label{kvs}
\xi_x = -\sin\phi \frac {\partial\ }{\partial \theta}-\cot \theta \cos\phi\frac{\partial\ }{\partial \phi}  \ \ \ \ \ \xi_y =\cos \phi \frac {\partial\ }{\partial \theta}-\cot \theta \sin\phi\frac{\partial\ }{\partial \phi} \ \ \ \ \ \xi_z = \frac{\partial\ }{\partial \phi}
\end{equation}
The lie bracket of these vectors yields
$
[\xi_i,\xi_j] = \epsilon_{ijk} \xi_k
$.
Though expressed in (\ref{kvs}) in a single chart on $S^2$, these vectors are in fact globally defined; precisely they can be extended to smooth  vector fields defined everywhere on $S^2$. This is obvious as the infinitesimal generators of  rotations about the origin induce a  Killing vector field everywhere on the 2-sphere embedded in ${\mathbb R}^3$. Hence $S^2$ with the round metric is globally spherically symmetric. 

${\mathbb R}P^2$ is a  second manifold  that has a globally spherically symmetric metric. It can be constructed from $S^2$ by identification of antipodal points; ${\mathbb R}P^2 = S^2/\sim$ where the identification map $P$ is  $(x,y,z) \sim (-x,-y,-z)$ restricted to the unit 2-sphere.  This space only locally embeds in ${\mathbb R}^3$. However, it is still globally invariant under $SO(3)$; $3\times 3$ matrices generating rotations commute with $P$. 

Any globally spherically symmetric
space must also be locally spherically symmetric. However, 
converse is not true. A simple example  is given by
 periodic identifications on  ${\mathbb R}^3$ with flat metric;
the resulting space is a 3-torus, $T^3$. The periodic translations leave
 ${\mathbb R}^3$ invariant. Therefore, $T^3$ is locally spherically symmetric; pick a neighborhood consisting of an open ball around any point $p$ of radius smaller than that of the identification period. As the space is flat, the metric is clearly rotationally invariant about $p$. However, it is not globally invariant under $SO(3)$. The  rotational Killing vector fields defined in the neighborhood around $p$ cannot be smoothly extended to all of $T^3$ as they become multiply defined 
under the periodic identification. Hence they do not correspond to global Killing vector fields and $T^3$, though locally spherically symmetric, is not globally spherically symmetric.

We now give a proof of Birkhoff's theorem for vacuum spherically symmetric spacetimes with or without cosmological constant. 
\begin{thm}\label{CosBirk} The only locally
spherically symmetric solutions to $R_{ab}=\Lambda g_{ab}$
are locally isometric either to one of the Schwarzschild-de Sitter (anti-de Sitter)
family of solutions or to Nariai spacetime.  Furthermore, these solutions are real
analytic in each local coordinate chart.
\end{thm} 

\begin{proof}

As the spacetime is locally spherically symmetric, $SO(3)$ acts in a neighborhood of any point such that its orbits are locally two dimensional spacelike surfaces. Choose coordinates in this neighborhood such that these surfaces are parameterized by $\theta, \phi$ with metric of the form (\ref{spherical}). Then the general form of the spacetime metric can be written as
\begin{equation}\label{smetric}
ds^2 = -A(T,R) {dT^2}+ 2B(T,R)dTdR+C(T,R)dR^2 + 
Y^2\left( T,R \right) d\Omega _2^2\end{equation}
where the functions $A$ and $C$ satisfy the condition $AC+B^2>0$, i.e. the metric is lorentzian. This metric can be brought to the form
\begin{equation}\label{ssmetric}
ds^2 = F(u,v)du^2+ 2X(u,v)dudv + 
Y^2\left( u,v \right) d\Omega _2^2\end{equation}
through a coordinate transformation.\footnote{First note that (\ref{smetric}) can be rewritten as
$ds^2 = -A(dT+DdR)^2+ 2(AD+B)(dT+DdR)dR + 
Y^2 d\Omega _2^2$
where  $D$ is chosen such that $C-AD^2-2BD=0$. Now choose new coordinates $v=R$ and $u(T,R)$  such that $du = f(dT + D dR)$ is a total differential for appropriately chosen $f(T,R)$. Defining $F = -A/f^2$ and $X = (AD+B)/f$ yields the metric in the form (\ref{ssmetric}). Note that this metric is lorentzian for $X\neq 0$ and any $F$. }
The vacuum Einstein equations with cosmological constant for (\ref{ssmetric}) are equivalent to

\begin{align}%G^u_v&&
-\frac{2}{X^2Y} \left(-\partial_v X\partial_v Y + X \partial_v^2Y\right) &=0   \label{ssequation1}\\
	%G^v_v:&&
	 -\frac{1}{X^2Y^2}\left( X^2 + Y\partial_vF\partial_vY + F(\partial_vY)^2 - 2X\partial_uY\partial_vY -2XY\partial_u\partial_vY\right)&=- \Lambda  \label{ssequation2}\\
%G^\theta_\theta-\frac{F}{2X}G^u_v:&&
 \frac{1}{2X^3Y}(Y\partial_vF\partial_vX - 2X\partial_vF\partial_v Y - XY\partial_v^2F - 2Y\partial_uX\partial_v X &\nonumber\\
 %&&
 + 2 XY \partial_u\partial_v X + 4 X^2 \partial_u\partial_v Y)&= -\Lambda   \label{ssequation3}\\
	%R_{uu}=
	% \frac{1}{2X^3Y} (-F Y\partial_vF\partial_v X+ 2FX\partial_vF\partial_vY+ F X Y \partial_v^2 F + 2 X^2 \partial_u F\partial_vY
	% +2 F Y \partial_uX\partial_vX &\\ -4FX \partial_uX \partial_vY-2X^2 \partial_v F \partial_uY + 4 X^2 \partial_vX\partial_vY
	% -2FXY\partial_u\partial_vX - 4X^3\partial_u^2Y) &=
	%\Lambda F 
	%R_{uu}+ F G^\theta_\theta:&&
	 \frac{1}{X^3Y} ( X^2 \partial_u F\partial_vY + 2 X^2 \partial_u\partial_v Y
	 -2FX \partial_uX \partial_vY-X^2 \partial_v F \partial_uY&\nonumber\\
	 %&&
	 + 2 X^2 \partial_vX\partial_vY
	- 2X^3\partial_u^2Y) &=0\ .\label{ssequation4}
	\end{align}
Observe that  (\ref{ssequation1}) implies that $\partial_v(\frac{\partial_vY}{X})=0$.  Thus either $\partial_v Y = 0$ or $X =\xi(u)\partial_v Y$ where $\xi(u)$ is a function of $u$ only and is nonvanishing. 

 If $\partial_v Y = 0$, then  (\ref{ssequation2}) immediately implies that $Y=\frac{1}{\sqrt{\Lambda}}$. Hence this case only occurs if
$\Lambda>0$.
The metric (\ref{ssmetric}) is then the direct sum of a 2-dimensional lorentzian metric
and that of a round sphere. Next, note that $X$ can be set to $\pm 1$ by a further coordinate transformation.\footnote{Choose $U=u$ and $V(u,v)$ such that
$\pm dV = D du +Bdv$ where $D(u,v)$ is chosen such that $\partial_v D = \partial_uB$. The metric then takes the form
$ds^2 = (F-2D)dU^2 \pm 2 dU dV +  \frac{1}{\Lambda} d\Omega^2_2$
which is of the  form (\ref{ssmetric}).} Equation (\ref{ssequation3}) now implies that
$\partial_v^2F = 2\Lambda$, with solution $F= \Lambda v^2-1$ in a suitable coordinate choice. Thus the solution for this case is
\begin{equation}\label{nariaiss}
ds^2 = (\Lambda v^2 - 1) du^2 \pm 2dudv + \frac 1{\Lambda} d\Omega^2_2 
\end{equation}
the Nariai metric in Vaidya form.
Note that $\partial_u$ is a local Killing vector field for this metric.

If $X =\xi(u){ \partial_v Y}$, then (\ref{ssmetric}) takes the form
$ds^2 = (F-2\partial_uY) du^2 + 2\xi  du dY + Y^2 d\Omega^2_2$ upon noting $dY = \partial_uY du + \partial_vY dv$. A simple coordinate transformation
%\footnote{Namely let $v=Y$, $U(u)$ such that $dU =\pm \xi du$, then redefine $F$ and $U$.} 
yields a metric of the form
\begin{equation}\label{prevaidya}
ds^2 = F du^2 \pm 2 du dv + v^2 d\Omega^2_2\ .
\end{equation}
Now (\ref{ssequation2}) becomes $ \frac 1{v^2}(1 + v\partial_vF + F) = \Lambda$ with solution $F = \frac{\Lambda}{3} v^2 +\frac{2M(u)}{v} -1$ where $M(u)$ is a function of $u$ only.  At this point, (\ref{prevaidya}) is the Vaidya metric. Finally, (\ref{ssequation4}) implies that
$\partial_u M = 0$; hence $M$ is a constant. Furthermore, as $M$ is  an arbitrary constant of integration, it can be positive or negative.
The solution for this case is thus
\begin{equation}\label{vaidyass}
ds^2 = (\frac{\Lambda }{3}v^2+\frac {2M}{v} - 1) du^2 \pm 2dudv +v^2d\Omega^2_2 
\end{equation}
the Schwarzschild-de Sitter (Schwarzschild-anti-de Sitter) metric in Vaidya form.
Again $\partial_u$ is a local Killing vector field for this metric.

Analyticity of these solutions in a coordinate chart follows immediately from their forms, (\ref{nariaiss}) and (\ref{vaidyass}). 
If in a given chart, a solution is
in the Schwarzschild-de Sitter (anti-de Sitter) family with mass parameter $M$, then
it cannot have  a different mass parameter in any overlapping chart  due to analyticity. In addition, in the case of positive cosmological constant, it  cannot become a Nariai solution in any  overlapping chart if it is Schwarzschild-de Sitter in the first.
 It follows that the only locally spherically symmetric solutions to
the vacuum Einstein equations with cosmological constant are everywhere locally isometric to the Schwarzschild-de Sitter (anti-de Sitter) solution with mass parameter $M$ or the Nariai solution if $\Lambda>0$. These solutions all have an additional local Killing vector field of translational form.

\end{proof}

The metric (\ref{vaidyass}) can be brought into
 familiar form. 
If $\Lambda v^2 + \frac {2M}{v} -1<0$ in a given neighborhood, then $\partial_u$ is a timelike Killing vector field. A  coordinate transformation\footnote{Specifically, $r=v$ and  $t(u,v)$ such that $dt=du\pm \frac {dv}{\frac{\Lambda}{3} v^2+\frac{2M}{v}-1}$.}  yields  the Schwarzschild-de Sitter (Schwarzschild-anti-de Sitter) solution in the region exterior to the black hole,
\begin{equation}
ds^2 = -\left( 1- \frac {2M}{r}- \frac \Lambda{3} r^2 \right) dt^2+\frac{dr^2}{1- \frac {2M}{r}- \frac \Lambda{3} r^2} + r^2d\Omega _2^2 \ . \label{ssmetricoutside}
\end{equation}
 This solution is  manifestly locally static.

If $\Lambda v^2 + \frac {2M}{v} -1>0$ in a given neighborhood, then $\partial_u$ is a spacelike Killing vector field. A coordinate transformation\footnote{Specifically, $t=v$ and $r(u,v)$ such that $dr=du\pm \frac {dv}{\frac{\Lambda}{3} v^2+\frac{2M}{v}-1}$.}  yields
\begin{equation}
ds^2 = -\frac{dt^2}{\left( \frac \Lambda{3} t^2 + \frac {2M}{t} - 1\right) } + \left(\frac \Lambda{3} t^2 + \frac {2M}{t} - 1\right)dr^2 + t^2d\Omega _2^2 \ , \label{ssmetricinside}
\end{equation}
 the interior solution of the Schwarzschild-de Sitter (Schwarzschild-anti-de Sitter) spacetime. It is manifestly not locally static; rather it exhibits a translational spacelike Killing vector field $\partial_r$. At a fixed time, $t$, the spatial metric is just the product metric of the interval and a round 2-sphere. Hence this metric is of the form of a Kasner solution, a cosmological solution also sometimes called the Kantowski-Sachs solution. 
 
 Finally, the exceptional case of the Nariai solution (\ref{nariaiss}) can  also be brought into a familiar form through a further coordinate transformation;\footnote{Specifically, $r=v$ and $t(u,v)$ such that $dt=du\pm \frac {dv}{\Lambda v^2-1}$. }
\begin{equation}
ds^2 = -(1 - \Lambda r^2)dt^2  + \frac{dr^2} {(1 - \Lambda r^2) }+ 
\frac 1{\Lambda} d\Omega _2^2\  . \label{nariaimetric}
\end{equation}
This solution is manifestly locally static. The transformation to the form \begin{equation}
ds^2 = -\frac{dt^2}{( \Lambda t^2-1)}  + {( \Lambda t^2-1)} {dr^2} + 
\frac 1{\Lambda} d\Omega _2^2\   \label{nariaimetric2}
\end{equation}
can also be made; this form of the metric is not explicitly locally static, but in fact also has a timelike Killing vector due to the $SO(2,1)$ symmetry of the 2-dimensional de-Sitter metric. 

It is key to observe that the values of $M$ and $\Lambda$ determine whether or not $\Lambda v^2 + \frac {2M}{v} -1$ is positive or negative. This determines if the metric (\ref{vaidyass}) can be cast into form (\ref{ssmetricoutside}) or form (\ref{ssmetricinside}) and, simultaneously, if the local Killing vector field is timelike or spacelike.  

\begin{figure}
\centering
\includegraphics{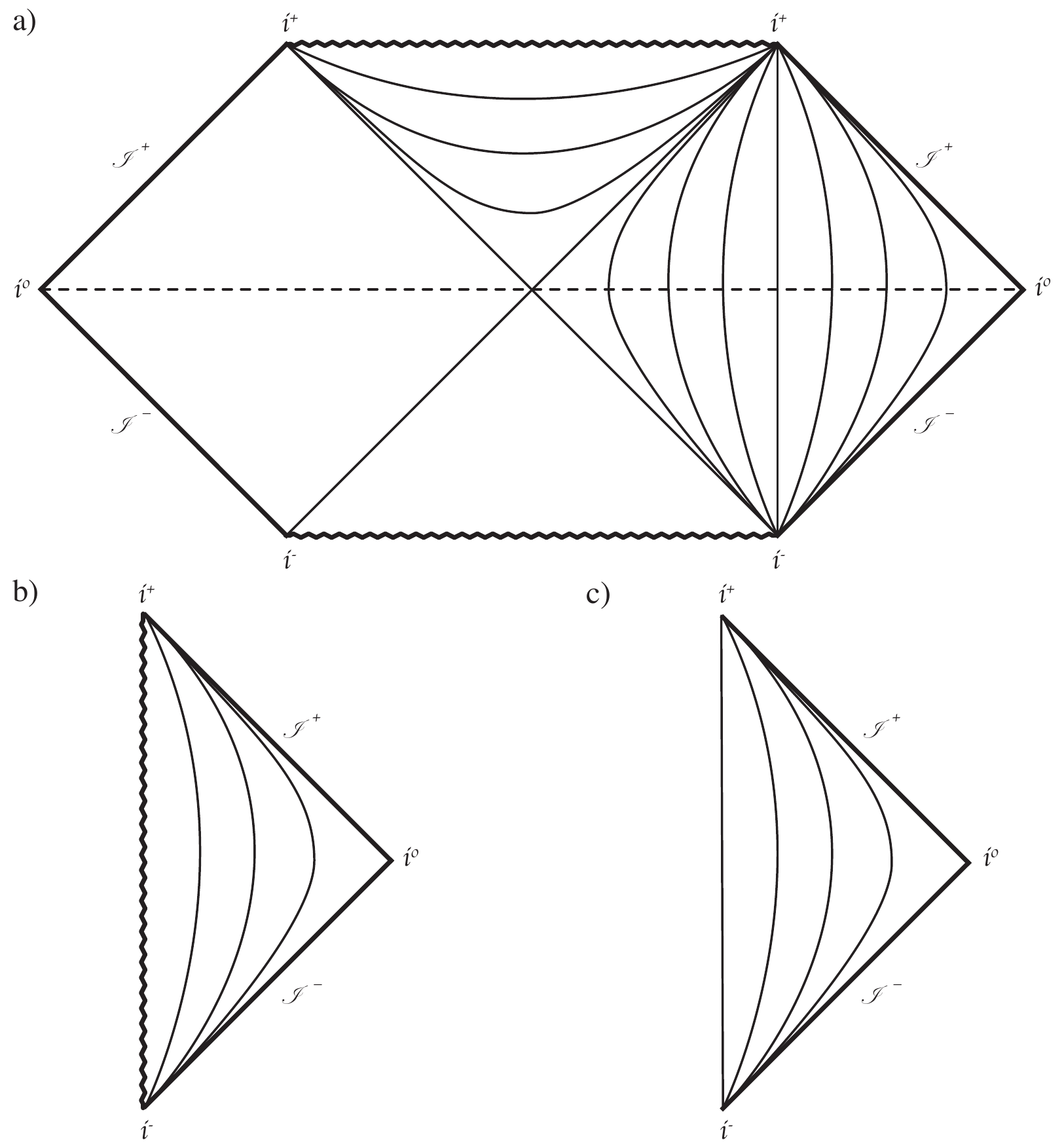}
\caption{The Penrose diagrams for the maximal analytic extension of Schwarzschild spacetime. Figure a) is the positive mass case.  The dotted line is the Cauchy surface with zero extrinsic curvature; its topology  is ${\mathbb R}\times S^2$. The negative mass case is in Figure b); the spacetime contains a naked singularity at $r=0$. Figure c) is the case of zero mass, Minkowski spacetime; the singularity seen in the negative mass case is now a coordinate singularity. Integral curves of the Killing vector field have been sketched in the exterior and black hole regions in Figure a) and in all regions in Figures b) and c). }
\label{fig:Schwarzschild}
\end{figure}

\begin{figure}
\centering
\includegraphics{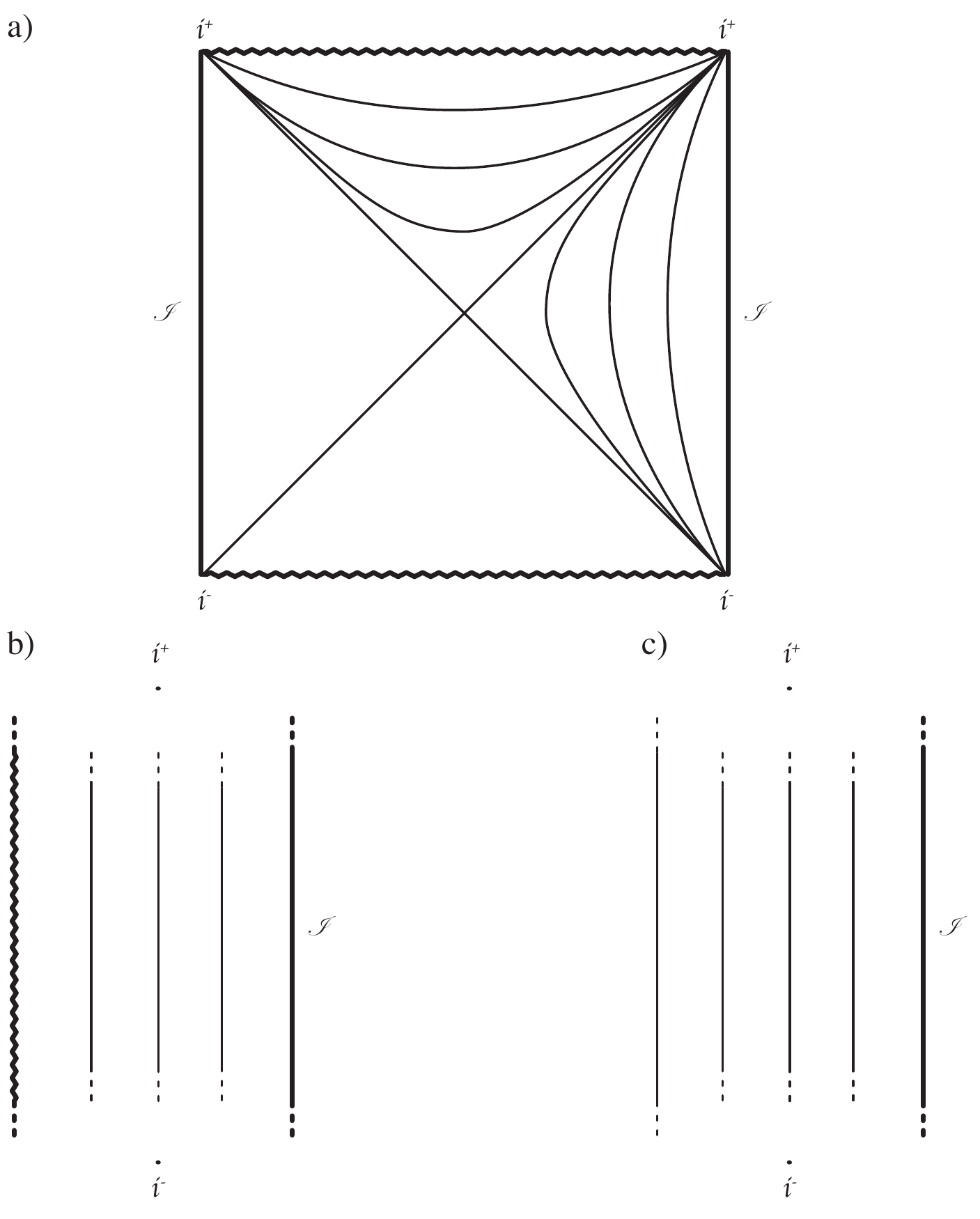}
\caption{The Penrose diagrams for the maximal analytic extension of Schwarzschild-anti-de Sitter spacetime. Figure a) is the positive mass case. Integral curves of the Killing vector field have been sketched in the exterior and black hole regions. Figures b) and c) are the negative mass and zero mass cases respectively. These diagrams extend infinitely. Note that  $\scri$ is timelike and that $i^+$ and $i^-$ are disconnected points. The negative mass case contains a naked singularity; this naked singularity becomes a coordinate singularity in the zero mass case, anti-de Sitter spacetime. Integral curves of the Killing vector field, sketched in both  Figures b) and c), also extend infinitely as indicated. }
\label{fig:AdS}
\end{figure}

When $\Lambda \leq 0$, 
if $M>0$, then  (\ref{vaidyass}) can be cast into the form of   (\ref{ssmetricoutside}) for the range of $v$ where $\partial_u$ is timelike and the form of (\ref{ssmetricinside}) where it is spacelike. Both of these possibilities occur for  different ranges of $v$ for $v>0$.\footnote{The metric  (\ref{vaidyass}) has a curvature singularity at $v=0$; hence it suffices to consider  $v<0$ and $v>0$. However, the case of $v<0$ is diffeomorphic to that of  $v>0$ with $M\to-M$. Thus it suffices to consider $v>0$ if both positive and negative $M$ are considered.} The maximal analytic extensions of Schwarzschild spacetime and Schwarzschild-anti-de Sitter spacetime are canonical examples of this behavior for $\Lambda=0$ and $\Lambda<0$ respectively; their conformal compactifications have the Penrose diagrams  in Figures \ref{fig:Schwarzschild} and \ref{fig:AdS}. 

If $M\leq 0$, then the Killing vector $\partial_u$ is timelike for all $v>0$ and only the form (\ref{ssmetricoutside}) is possible. Classic examples of this behavior  is given by the negative mass Schwarzschild spacetime and Minkowski spacetime whose conformal compactifications have the Penrose diagrams in Figure \ref{fig:Schwarzschild}. The corresponding Penrose diagrams for the $\Lambda<0$ case are given in Figure \ref{fig:AdS}. These solutions are also allowed by Birkhoff's theorem; negative mass solutions can only be eliminated by imposing additional conditions appropriate in a specific application.

\begin{figure}
\centering
\includegraphics{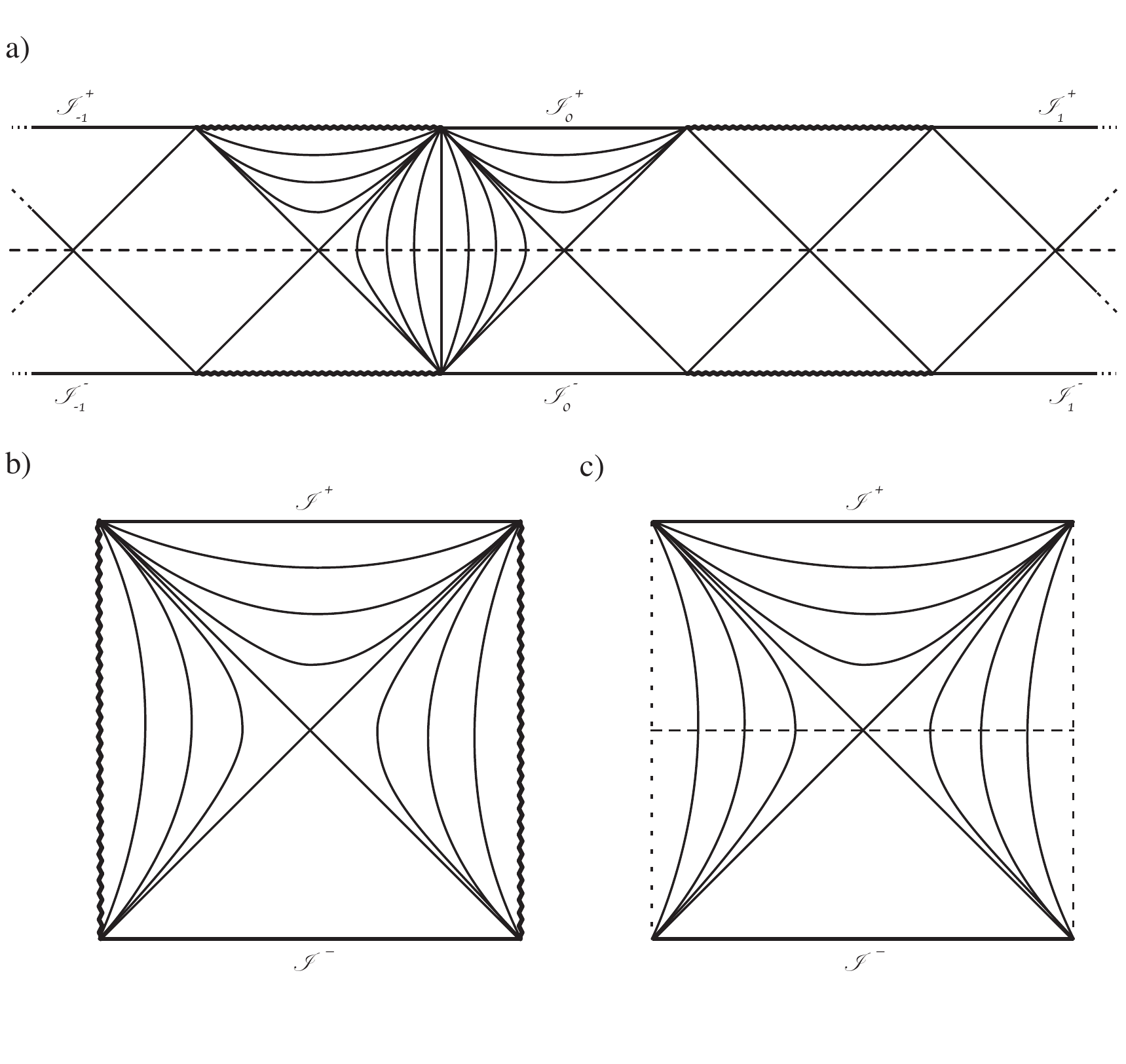}
\caption{The Penrose diagrams for the maximal analytic extension of Schwarzschild-de Sitter spacetime. Figure a) is that for $M>0$ and $0<\Lambda<\frac{1}{9M^2}$. The diagram repeats infinitely to both left and right. Integral curves of the Killing vector field have been sketched in a representative black hole and exterior region.  The dotted line is the Cauchy surface with zero extrinsic curvature; its topology is ${\mathbb R}\times S^2$. The negative mass case for any $\Lambda>0$ is in Figure b); the spacetime contains  naked singularities. Figure c) is the zero mass case, de Sitter spacetime; the singularities of the negative mass case are now coordinate singularities. Note that the topology of its Cauchy surface with zero extrinsic curvature is $S^3$. Integral curves of the Killing vector field have been sketched in both Figures b) and c).}
\label{fig:SdS}
\end{figure}

The situation becomes more complicated when $\Lambda> 0$ and $M>0$; whether or not there exists any neighborhood with a timelike Killing vector now depends on the value of both parameters. For 
$0<\Lambda<\frac{1}{9M^2}$, there are two real positive roots of $\Lambda v^2 + \frac {2M}{v} -1$,
\begin{align*}
r_h&=3MlR\left(1-\sqrt{1-\frac{1}{lR^3}}\right)\\
r_c&=3MlR\left(1+\sqrt{1-\frac{1}{lR^3}}\right)\ 
\end{align*}
where $l = (9M^2\Lambda)^{-1/2}$ and $R=\cos\left(\frac{1}{3}\cos^{-1}\left(\frac{1}{l}\right)\right)$ evaluated using the smallest  angular value of $\cos^{-1}$. 
The Killing vector $\partial_u$ is timelike for $r_h<v<r_c$ and consequently (\ref{vaidyass}) can be reexpressed 
in the form  (\ref{ssmetricoutside}). For $v<r_h$ and $v>r_c$, $\partial_u$ is spacelike; hence (\ref{vaidyass}) takes the form
(\ref{ssmetricinside}) in these regions. A further coordinate transformation to Kruskal form \cite{Gibbons:1977mu,Lake:1977ui} or alternately Israel form \cite{Lake:2005bf} results in the maximal analytic extension of the Schwarzschild-de Sitter spacetime, whose conformal compactification is illustrated in the Penrose diagram in
Figure \ref{fig:SdS}.

\begin{figure}
  \centering
\includegraphics{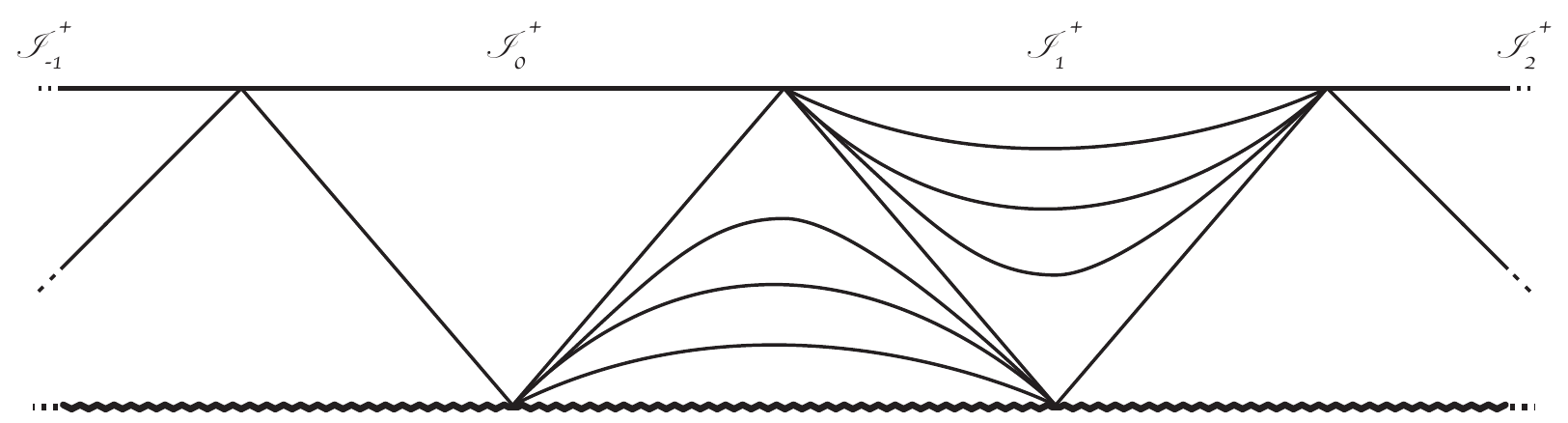}
\caption{The Penrose diagram for the maximal analytic extension of the Schwarzschild-de Sitter spacetime with  $M> 0$ and $\Lambda=\frac{1}{9M^2}$. The diagram repeats infinitely to both left and right. Integral curves of the Killing vector field have been sketched in a representative white hole and asymptotically de Sitter region. This diagram corresponds to the white hole case; the black hole case is given by the reflection of this diagram about the horizontal axis. Both cases are allowed as the spacetime does not have a preferred time orientation. }
\label{fig:XSdS}
\end{figure}

 When $\Lambda = \frac{1}{9M^2}$, a case sometimes termed the extremal Schwarzschild-de Sitter spacetime, the two roots coincide, $r_h=r_c = 3M$. Consequently, $\partial_u$ is spacelike for $0<v<\infty$ except at $v= 3M$ where $\partial_u$ is null. Thus this spacetime has no locally static region. However, it is still a well defined lorentzian spacetime and can be cast into the form (\ref{ssmetricinside}). A further coordinate transformation \cite{Lake:1977ui,Lake:2005bf,Podolsky:1999ts} yields its maximal analytic extension with the Penrose diagram illustrated in Figure \ref{fig:XSdS}.

When $\Lambda > \frac{1}{9M^2}$, there are no real positive roots. Consequently, $\partial_u$ is always spacelike and again the metric can only be cast into the form  (\ref{ssmetricinside}). Thus the over-extremal Schwarzschild-de Sitter spacetime has no static region. However, in contrast to the extremal case, this spacetime also has  no Killing horizons. In conclusion, when $\Lambda \geq \frac{1}{9M^2}$ and $M>0$, the maximal analytic extension of the locally spherically symmetric solution to the Einstein equations is never locally static.

If $\Lambda>0$ and $M\le 0$, then $\partial_u$ is  timelike for  $0<v< -6MlR$ and spacelike for $v>-6MlR$. Therefore, (\ref{vaidyass}) can be reexpressed in the form of (\ref{ssmetricoutside})  or  (\ref{ssmetricinside})  in each region respectively. The Penrose diagrams for this case are given in Figure \ref{fig:SdS}.

\begin{figure}
\centering
\includegraphics{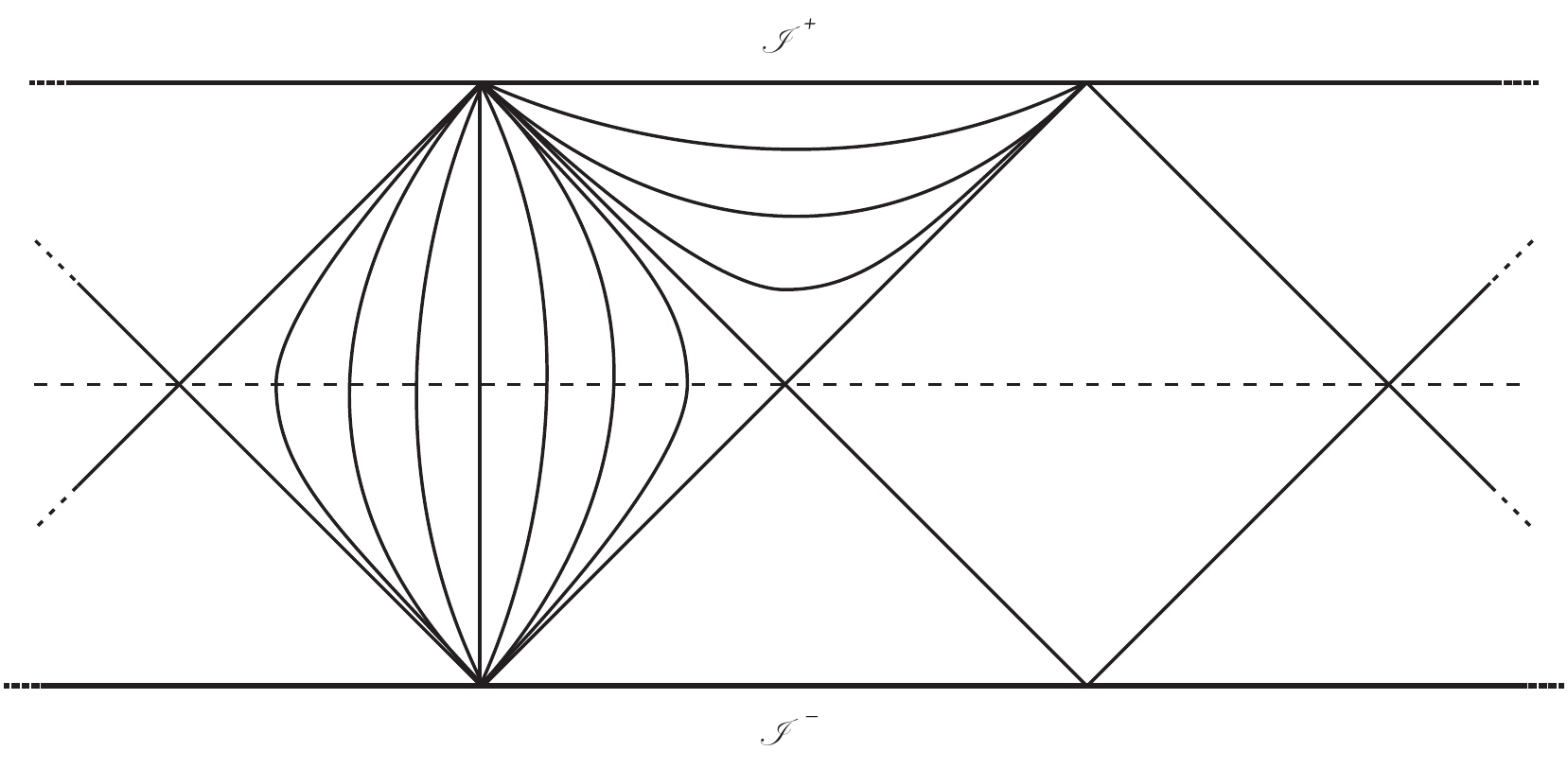}
\caption{The Penrose diagram for the maximal analytic extension of the Nariai spacetime.  The diagram repeats infinitely to both left and right. Integral curves of the Killing vector field for one possible choice have been sketched in representative regions. The dotted line is the zero extrinsic curvature Cauchy surface with topology ${\mathbb R}\times S^2$. }
\label{fig:Nariai}
\end{figure}

Finally, the Penrose diagram for the maximal analytic extension of the Nariai spacetime is given in Figure \ref{fig:Nariai}. It differs from the Penrose diagram of 4-dimensional de Sitter spacetime in that it extends infinitely, reflecting the fact that its Cauchy surface has topology ${\mathbb R}\times S^2$.

Although the proof presented here has certain similarities to traditional proofs such as that of \cite{hebirk},  it does not rely on a foliation of the neighborhood by surfaces of constant $Y$. Rather, the equations of motion are used to construct the solution from a choice of coordinates resulting in a non-diagonal metric of simple form. This anzatz also yields a local chart that extends, if desired,  across  black and white hole horizons and, if $\Lambda >0$, cosmological horizons. Note that the proof explicitly exhibits the fact that the metric in every local chart has a fourth Killing vector field of translational form. However, the metric in this local chart is not necessarily static as the Killing vector field is not necessarily timelike. Furthermore, the maximal analytic extension of the Schwarzschild-de Sitter case  need not exhibit a static region. Hence the common belief that Birkhoff's theorem implies staticity is false for the case of positive cosmological constant. 
The same issue is likely to hold in generalizations of Birkhoff's theorem to other theories of gravity  with spherically symmetric solutions similar to the Schwarzschild-de Sitter solutions. In any case, the correct view of Birkhoff's theorem and its generalizations is that they are local uniqueness theorems that additionally imply that these solutions have an additional local Killing vector field. This distinction may be especially relevant in cosmological settings in which alternative theories of gravity are of current interest.

\acknowledgments The work was supported by NSERC. In addition, the authors would like to
thank the Perimeter Institute for its hospitality.


\end{document}